\documentclass[12pt]{article}
\usepackage{graphics,epsfig}
\renewcommand{\theequation}{\arabic{section}.\arabic{equation}}
\newcounter{saveeqn}
\newcommand{\alpheqn}{\setcounter{saveeqn}{\value{equation}}%

\setcounter{equation}{0}\addtocounter{saveeqn}{1}%

\renewcommand{\theequation}{\arabic{section}.\arabic{saveeqn}.
\alph{equation}}}
\newcommand{\reseteqn}{\setcounter{equation}{\value{saveeqn}}%
\renewcommand{\theequation}{\arabic{section}.\arabic{equation}}}

\newcommand{\iw}{\int _y^x\hskip-17pt {\bf w}\hskip2pt}

\begin{document}

\begin{titlepage}\hfill HD-THEP-01-17

\hfill Eur. Phys. J. C, in print\\[4ex]

\begin{center}{\Large \bf A Gauge and Lorentz covariant Approximation for
the Quark Propagator in an arbitrary Gluon Field} \\[5 ex]

{\bf Dieter Gromes}\\[3 ex]Institut f\"ur
Theoretische Physik der Universit\"at Heidelberg\\ Philosophenweg 16,
D-69120 Heidelberg \\ E - mail: d.gromes@thphys.uni-heidelberg.de \\
 \end{center} \vspace{2cm}

{\bf Abstract:} We decompose the quark propagator in the
presence of an arbitrary gluon field with respect to
a set of Dirac matrices. The four-dimensional integrals which arise in first order perturbation theory are rewritten as line-integrals along certain field lines, together with a weighted integration over the various field lines. It is then easy to transform
the propagator into a form involving path ordered exponentials. The
resulting expression is non-perturbative and has the correct behavior
under Lorentz transformations, gauge transformations and charge
conjugation. Furthermore it coincides with the exact propagator in
first order of the coupling $g$. No expansion with respect to the
inverse quark mass is involved, the expression can even be used for
vanishing mass. For large mass the field lines concentrate near the straight line connection and simple results can be obtained immediately.

 \vfill \centerline{March 2001}

\end{titlepage}

\section{The quark propagator }

The quark propagator $S(x,y;A)$ for a quark of mass $m$ in the
presence of a gluon field $A^\mu $ plays an important role in many
investigations of quantum chromodynamics. It appears, e.g., if one
considers the quark four-point Green function, the basis of all
modern investigations on quark-antiquark interactions, and integrates
over the quark fields. It is defined by \footnote{Our formulae are
given in Minkowsi space, we use Bjorken Drell conventions \cite{BD},
and the field tensor is defined by $F_{\mu \nu } = \partial _\mu A_\nu
-\partial _\nu A_\mu -ig[A_\mu ,A_\nu ]$.}

\begin{equation} S_{\alpha \beta }(x,y;A) = -i <0|T(\psi _\alpha
(x)\bar{\psi }_\beta (y))|0>.  \end{equation}
The spinor indices $\alpha ,\beta $ will be dropped in the following.
We recall the relevant properties of the propagator. With the
covariant derivative $D_\mu (x)$ which acts on operators to the right, and $\stackrel{\leftarrow}{D}^*_\mu (y)$ which acts on operators to the left, defined by

\begin{eqnarray} D_\mu (x) & = & \frac{\partial }{\partial x^\mu }
 -igA_\mu (x), \nonumber\\
\stackrel{\leftarrow}{ D}^*_\mu (y) & = &
\frac{\stackrel{\leftarrow}{\partial }}{\partial y^\mu }+ igA_\mu (y),
\end{eqnarray}
the field equations give

\begin{eqnarray} [i\gamma ^\mu D_\mu (x) -m]S(x,y;A) & = & \delta
^{(4)}(x-y), \nonumber\\
 S(x,y;A)[-i\gamma ^\mu \stackrel{\leftarrow}{ D}^*_\mu (y) -m]
& = & \delta ^{(4)}(x-y). \end{eqnarray}
These equations may be reformulated as integral equations:

\begin{eqnarray} S(x,y;A) & = & S_0(x-y) - g \int S(x,z;A)\gamma ^\mu
A_\mu (z) S_0(z-y)d^4z \nonumber\\
& = & S_0(x-y) - g \int S_0(x-z) \gamma ^\mu A_\mu (z) S(z,y;A)d^4z,
 \end{eqnarray}
with $S_0$ the free propagator. From charge conjugation one has

\begin{equation} S(x,y;A) = \gamma ^2\gamma ^0S^T (y,x;-A^T)
\gamma ^2\gamma ^0. \end{equation}
Finally, under a gauge transformation $\psi \rightarrow \psi ' =
e^{i\Theta }\psi ,\bar{\psi } \rightarrow \bar{\psi }' = \bar{\psi }
e^{-i\Theta }, A_\mu \rightarrow {A'}_\mu = e^{i\Theta} A_\mu
e^{-i\Theta} - (i/g)(\partial _\mu e^{i\Theta })e^{-i\Theta }$, the propagator transforms
as

\begin{equation} S(x,y;A) \rightarrow S'(x,y;A) = e^{i\Theta (x)}
S(x,y;A) e^{-i\Theta (y)}.  \end{equation}
An exact solution for $S(x,y;A)$ for an arbitrary gluon field $A_\mu $
is not available. However, it would be
highly desirable to have an approximation which respects the
fundamental properties of the propagator, in particular the correct
transformation under Lorentz transformations and under gauge
transformations.

The two well known approximations, perurbation theory and the static approximation, either violate gauge covariance or Lorentz covariance: 

\begin{description}
\item[Perturbation theory:] Iteration of eq. (1.4) gives the
perturbation series

\begin{equation} S(x,y;A) = S_0(x-y) - g \int S_0(x-z)\gamma ^\mu
A_\mu (z) S_0(z-y)d^4z + \cdots . \end{equation}

Any finite order of the perturbation series gives the correct behavior
under Lorentz transformations and under charge conjugation. It will,
however, never be able to describe central features of QCD like
confinement. In particular, it clearly violates gauge invariance; no
truncation of the perturbation series (1.7) has the correct
transformation property (1.6).

\item[Static approximation:] Following the pioneering work of Brown,
Weisberger \cite{BW} and Eichten, Feinberg \cite{EF}, one neglects the
spatial part $i\gamma ^mD_m(x)$ in eq. (1.3). The partial differential
equation then becomes an ordinary differential equation which can be
solved in closed form. This leads to the static approximation

\begin{eqnarray} S_{stat}(x,y;A) & = & -i \left\{ \Theta
(x_0-y_0)\frac{1+\gamma ^0}{2} + \Theta (y_0-x_0) \frac{1-\gamma ^0}{2
}\right\}
\nonumber\\ & & \delta ^{(3)}({\bf x} - {\bf y}) e^{-im_|x_0 -y_0|}
P \exp\{ig \int _y^x A_0 (z)dz^0 \}. \end{eqnarray}

The path in the line integral is the straight line from $y$ to $x$,
and the path ordering orders $A_0 (x)$ to the left, $\; \cdots, \; A_0 (y)$ to
the right.

The neglected spatial term $i\gamma ^mD_m(x)$ in (1.3) can
subsequently be taken into account as perturbation. This approach has
been extremely succesful (see e.g. the review \cite{rev}). It can be
easily generalized to quarks moving with four velocity $v^\mu $, and
thus is the direct predecessor of heavy quark effective theory.
Succesful combination with perturbation theory has also been made more
recently (see \cite{AA} and references therein). For quarks moving with high momentum a related formula can be derived from the eikonal approximation \cite{nach}.

The static approximation and it's generalizations are non-perturbative and have the correct
behavior under gauge transformations and under charge conjugation.
However, they drastically violate Lorentz invariance. Therefore the static approximation is useful for heavy quarks only. It needs quite an effort to
recover the relations following from the original Lorentz invariance subsequently \cite{rev,BGV}. Finally the static approximation does not coincide
with the exact propagator even in the trivial case of vanishing gluon
field. \end{description}

It is obvious that a gauge covariant propagator should contain path ordered exponentials as in (1.8). Using just the path along the straight line between $x$ and $y$ would give a Lorentz covariant result, but such a procedure would be far too simple. It would involve the vector potential only along the straight line connection and nowhere else, which is clearly unphysical. We prefer to  proceed systematically by rewriting perturbation theory in a suitable way. It can then easily be transformed into a gauge covariant expression by simple exponentiation, while keeping the correctness of perturbation theory in the relevant order. The non-perturbative approximation for the quark propagator obtained in this way has the following properties:

\begin{itemize}
\item Correct behavior under Lorentz transformations.
\item Correct behavior under gauge transformations.
\item Correct behavior  under charge conjugation.
\item Agreement with perturbation theory in first order of the coupling.
\end{itemize}
The representation is a weighted superposition of path ordered
exponentials between $x$ and $y$ along well defined field lines. We don't need to assume that the quark mass is large, we could even put it equal to zero.
This opens perspectives to applications which were hard to attack
previously.

The paper is organized as follows:

In sect. 2 we decompose the propagator with respect to Dirac matrices,
and write the formula of first order perturbation theory in a form
which is convenient for the following. In sect. 3 we rewrite the four-dimensional space-time integrals which arise in perturbation theory as weighted superpositions of line integrals over certain field lines which all run from $x$ to $y$. From this form one can simply derive a representation in terms of superpositions of path ordered exponentials. This representation coincides with perturbation theory up to order $g$ and has the correct behavior under
gauge transformations. In sect. 4 we evaluate the weight function explicitly and show a plot of the field lines for different masses. We discuss the limit of large mass $m$, and give some first simple applications. Actual applications will be given in forthcoming papers.

\setcounter{equation}{0}\addtocounter{saveeqn}{1}%
\section{A useful form of first order perturbation theory}

We start with the first order approximation (1.7), and
express the free propagators $S_0$ by the free scalar propagator
$\Delta $ in the following way:

\begin{eqnarray} S_0(x-z) & = & \Delta(x-z)[-i\gamma _\nu
\stackrel{\leftarrow}{\partial }/\partial z_\nu  + m],\nonumber\\
S_0(z-y) & = & [i \gamma _\lambda \partial /\partial z_\lambda +m]
\Delta (z-y). \end{eqnarray}
The free scalar propagator satisfies the equation

\begin{equation} (\partial _\rho \partial ^\rho + m^2)\Delta(x) = - \delta ^{(4)}(x).
\end{equation}
In the following all differential operators in the integrand are
understood as differentiations with respect to the variable $z$.

Using the well known identities $\gamma _\nu \gamma _\mu =g_{\nu \mu
} - i\sigma _{\nu \mu }$, and $\gamma _\nu \gamma _\mu \gamma _\lambda
= g_{\nu \mu }\gamma _\lambda - g_{\nu \lambda }\gamma _\mu + g_{\mu
\lambda }\gamma _\nu + i\epsilon _{\nu \mu \lambda \kappa }\gamma
^5\gamma ^\kappa $, one can write the propagator in form of the
familiar decomposition

\begin{equation} S(x,y;A) = s + p\gamma ^5 + v^\mu \gamma _\mu + a^\mu
\gamma ^5\gamma _\mu + t^{\mu \nu }\sigma _{\mu \nu }.  \end{equation}
Before giving the expressions for $s,\cdots ,t^{\mu \nu }$ which result
in this way, it is convenient for later use to define a scalar field $u(z;x,y)$ and a vector field $u^\mu (z;x,y)$ by

\begin{equation} u(z;x,y) = \Delta (x-z) \Delta (z-y),\end{equation}
and (with $\stackrel{\leftrightarrow}{\partial }^\mu =
\stackrel{\rightarrow}{\partial }/\partial z_\mu -
\stackrel{\leftarrow}{\partial }/\partial z_\mu $)

\begin{equation} u^\mu (z;x,y) = - \Delta^{-1}(x-y) [\Delta (x-z)
\stackrel{\leftrightarrow}{\partial }^\mu \Delta (z-y)].\end{equation}
From (1.7), (2.1), and (2.3) we then obtain the following equations:

\begin{eqnarray} s & = & m \Delta (x-y) [1 + ig\int u^\mu (z;x,y)A_\mu
(z)d^4 z], \\
p & = & 0, \\
v^\mu & = & \frac{i}{2}\partial /\partial
x_\mu  \Delta (x-y) - \frac{i}{2} \Delta (x-y)
\stackrel{\leftarrow}{\partial }/\partial y_\mu \\
& & -g \int \left\{\Delta(x-z) [m^2g^{\mu \nu } +
\stackrel{\leftarrow}{\partial }^\nu \stackrel{\rightarrow
}{\partial}^\mu - \stackrel{\leftarrow}{\partial }_\lambda g^{\mu \nu
}\stackrel{\rightarrow }{\partial } ^\lambda +
\stackrel{\leftarrow}{\partial }^\mu \stackrel{\rightarrow }{\partial
}^\nu ] \Delta(z-y) \right\}A_\nu (z)d^4z, \nonumber\\
a^\mu & = & ig \epsilon ^{\mu \nu \lambda \kappa } \int
[\Delta (x-z) \stackrel{\leftarrow}{\partial }_\kappa
\stackrel{\rightarrow }{\partial }_\lambda \Delta (z-y)]A_ \nu (z)d^4
z, \\
t^{\mu \nu } & = & -\frac{gm}{2} \int \left\{\partial _\nu u(z;x,y) A_\mu (z) - (\mu \leftrightarrow \nu ) \right\} d^4z.  
\end{eqnarray}
We next transform $v^\mu $ in (2.8), we denote the four terms in the second line by $v_1^\mu +v_2^\mu +v_3^\mu +v_4^\mu $. In $v_2^\mu $ one can use
$\stackrel{\rightarrow }{\partial }^\mu _z = - \stackrel{\rightarrow
}{\partial }^\mu _y$, in $v_4^\mu $ correspondingly $\stackrel{\leftarrow
}{\partial }^\mu _z = - \stackrel{\leftarrow }{\partial }^\mu _x$.
The differentiations with respect to $y$ and $x$ can then be taken
outside of the integral. The term $v_1^\mu $ is split into two identical
contributions, in the first one we use eq. (2.2) for $\Delta
(x-z)$, in the second one for $\Delta (z-y)$. Next perform a partial
integration on one of the derivatives in the d'Alembert operator. The
terms where the differentiations act on the other $\Delta $ cancel
against the contribution $v_3^\mu $, and one has the intermediate result

\begin{eqnarray} v_1^\mu  + v_3^\mu  & = & \frac{g}{2} \Delta (x-y) \{A^\mu (x) +
A^\mu (y)\} \nonumber\\
& & -\frac{g}{2}  \int \partial _\lambda u(z;x,y)
[\partial ^\lambda A^\mu (z) - \partial ^\mu A^\lambda (z)]d^4z
\nonumber\\
& & - \frac{g}{2} \int \partial _\lambda u(z;x,y)
\partial ^\mu A^\lambda (z))d^4z . \end{eqnarray}
We subtraced and added the term $\partial ^\mu A^\lambda (z)$. In the
second line we can then replace $\partial ^\lambda A^\mu (z)- \partial
^\mu A^\lambda (z)$ by $F^{\lambda \mu }(z)$, which only introduces a
higher order error $O(g^2)$. Furthermore the differential operator $\partial _\lambda = \partial _\lambda ^z$ which acts on $u(z;x,y)$ can be replaced by $-(\partial _\lambda^x + \partial _\lambda ^y)$ and taken outside of the integral. In the third line we perform a
partial integration on $\partial ^\mu $, shift the differentiations
from the variable $z$ to $x$ and $y$, and take the
differentiations out of the integral. After these manipulations $v^\mu
$ can be expressed in a rather compact form if we replace partial derivatives by covariant derivatives which introduces a correction of order $g^2$ only:
\newpage

\begin{eqnarray} v^\mu & = & \frac{i}{2} D^\mu (x) \left\{ \Delta
(x-y) [1 + ig \int u^\nu (z;x,y) A_\nu (z) d^4z]\right\} \nonumber\\
& & - \frac{i}{2}  \left\{ \Delta (x-y) [1 + ig \int u^\nu (z;x,y)
A_\nu (z) d^4z]\right\}\stackrel{\leftarrow}{D}^{*\mu }(y)
\nonumber\\
& & - \frac{g}{2} D_\nu(x) \int u(z;x,y) F^{\mu \nu}(z) d^4 z
- \frac{g}{2}  \int u(z;x,y) F^{\mu \nu}(z) d^4 z \stackrel{\leftarrow}{D}^*_\nu(y).\end{eqnarray}
In $a^\mu $ we perform a partial integration with respect to
$\stackrel{\leftarrow}{\partial }_\kappa $ or $\stackrel{\rightarrow}{\partial }_\lambda $ , take half of the sum of both terms, antisymmetrize the
$\partial _\kappa A_\nu (z)$ or $\partial _\lambda A_\nu (z)$ in the integrand, and introduce the field strength tensor $F_{\kappa \nu }(z)$ or $F_{\lambda \nu }(z)$ as before. Shifting again the differentiation from $z$ to $x$ and $y$ we have in order $g$

\begin{equation} a^\mu = \frac{ig}{4} \epsilon ^{\mu \nu \lambda
\kappa } \left\{D_\lambda (x) \int u(z;x,y) F_{\nu \kappa }(z) d^4z
- \int u(z;x,y) F_{\nu \kappa }(z) d^4z \stackrel{\leftarrow}{D}_\lambda ^*(y)
\right\}. \end{equation}
Similar manipulations applied to $t^{\mu \nu }$ finally give

\begin{equation} t^{\mu \nu } = -\frac{gm}{2}  \int u(z;x,y)
F^{\mu \nu }(z)d^4z. \end{equation}
The decomposition (2.3), together with the formulae (2.6), (2.7),
(2.12), (2.13), (2.14) is now in a form which
allows to rewrite the four-dimensional integrations $d^4z$ as a
superposition of line integrals.

\setcounter{equation}{0}\addtocounter{saveeqn}{1}%
\section{Gauge covariant reformulation}

The vector field $u^\mu (z;x,y)$ defined in (2.5) satisfies the
fundamental equation

\begin{equation} \frac{\partial u^\mu (z;x,y)}{\partial z^\mu } =
\delta ^{(4)}(z-y) - \delta ^{(4)}(z-x), \end{equation}
which is a simple consequence of eq. (2.2). Therefore $u^\mu $ may be
interpreted as a four-dimensional velocity field of an incompressible
fluid with a point-like source at $y$ and a sink at $x$. The stream
lines $z^\mu (s,{\bf w})$, which all run from $y$ to $x$, are
defined by the characteristic equations

\begin{equation} \frac{dz^\mu (s,{\bf w})}{ds} = u^\mu
(z(s,{\bf w})). \end{equation}
Here $s$ is the parameter which describes the motion along the stream
line, while the three dimensional parameter set ${\bf w}$ characterizes the
various stream lines. To make $s$ unique, it is convenient to fix $s=0$ at the symmetrical point of the stream line which has equal distance to $x$ and $y$. The dependence on $x,y$ has been suppressed in
the notation. There is precisely one field line passing through every
space time point, except, of course, for the source points $x$ and
$y$. After having solved (3.2) there is a unique
correspondence between the 4-dimensdional space-time coordinates
$z^\mu $ and the parameters $s,{\bf w}$, i.e. $(z^0,z^1,z^2,z^3)
\Leftrightarrow (s,w^1,w^2,w^3)$.

We next write the four-dimensional integrals over $d^4z$ which appear in $s$ and the first two lines of $v^\mu $ (eqs (2.6) and (2.12)) as integrals
over $dsd^3w$, the Jacobian is called $\rho ({\bf w})$:

\begin{equation} \rho ({\bf w}) = \frac{\partial
(z^0,z^1,z^2,z^3)}{\partial (s,w^1,w^2,w^3)}.  \end{equation}
We have anticipated the crucial fact that $\rho $ does not depend on
the curve parameter $s$. This is a direct consequence of the
incompressibility of the flow, and easily proved from the following
geometrical argument. Take an infinitesimal four-dimensional box
in $(s,{\bf w })$-space with corners $(s_0,{\bf w})$ and $(s_0 +
ds,{\bf w} + d{\bf w})$. In $z$-space this corresponds to a an
infinitesimal region with a certain volume. Consider now the motion of
this volume along a field line from $s_0$ to $s_1$, keeping ${\bf
w},d{\bf w}$ and $ds$ constant. Because of the vanishing divergence
(3.1) outside the sources, the volume in $z$-space stays constant, the
volume in $(s,{\bf w})$-space stays constant anyhow by construction. This
demonstrates that the Jacobian (3.3) is indeed independent of $s$.

We can therefore write

\begin{eqnarray} \int u^\mu (z;x,y) A_\mu (z) d^4z & = & \int \rho
({\bf w})[\int _y^xA_\mu (z(s,{\bf w}))u^\mu (z(s,{\bf w})) ds] d^3w
\nonumber\\
& = & \int \rho ({\bf w })[\iw Adz ]d^3w.
\end{eqnarray}
Here $\iw Adz $ is a shorthand notation for the line integral of $A_\mu $ from $y$ to $x$ along the stream line characterized by the parameter ${\bf w}$.

The normalization of $\rho ({\bf w})$ is easily obtained from the
special case $A_\mu (z) = \partial _\mu \Theta (z)$, where both
sides of (3.4) can be immediately integrated. This gives

\begin{equation} \int \rho ({\bf w})d^3w = 1.  \end{equation}
We are now in the position to rewrite the scalar function $s$ in
(2.6):
\alpheqn
\begin{eqnarray} s & = & m\Delta (x-y) \int \rho ({\bf w})[1 + ig
 \iw Adz ] d^3w \\
& = & m\Delta (x-y) \int \rho ({\bf w})P \exp[ig \iw Adz ] d^3 w + O(g^2). 
\end{eqnarray}
\reseteqn
This was the essential step of the approach! Once having written the first order approximation in terms of line integrals, exponentiation allows to promote it into a non-perturbative gauge covariant expression. 

One may wonder about the justification of this step which led to a non-perturbative expression simply by exponentiation. It does not make sense to compare the two lines in (3.6), because (3.6.b) is gauge covariant while (3.6.a) is not. In fact one can easily see, e.g. for a suitable pure gauge, that (3.6.a) can be made as large as one likes. The correct question to be asked is whether (3.6.b) is a reasonable approximation to the exact propagator. To answer this question one only needs to check the quality of the approximation in {\it some special convenient} gauge. The manifest covariance of the expression then guarantees this quality for {\it any} gauge. A sufficient condition is the smallness of $g\iw Adz$ for all relevant stream lines, which justifies the validity of first order perturbation theory. Clearly one can always find a gauge where $A_\mu dz^\mu = 0$ on one special stream line, but it is not possible to do the same for two or more lines. For heavy quarks the situation is simple. We will show in sect. 4 that in this case all relevant paths lie near the straight line connection. If one chooses a gauge such that  $A_\mu (x-y)^\mu $ vanishes along this line, and if the variations of $A_\mu $ near this line are small, the approximation will be justified. In applications we have to perform an integration over the gauge field $A_\mu $ at the end, which involves the whole spectrum in momentum space. Because there are only two scales available, $m$ and $\Lambda _{QCD}$, one expects a good approximation if $m>>\Lambda _{QCD}$. 

For light quarks the lines spread out over the whole space, and the above argument cannot be applied. But in any case the step of exponentiation leading from (3.6.a) to (3.6.b) can be interpreted as partially including higher orders of the perturbation series, namely a minimum of those necessary to guarantee gauge covariance. There are good reasons to believe that gauge covariance is such a fundamental principle that it may indeed be used to transform perturbative expressions into  non-perturbative ones of physical relevance. 

The expressions in the first and second line of $v^\mu $ can be treated in exactly the same way. The integrals in the third line of $v^\mu $, as well as those in $a^\mu $ and $t^{\mu \nu }$ have a different form, but the structure of all of them is identical. They have a factor $g$ in front, therefore one can introduce path ordered exponentials without changing the result in order $g$. To each $z\neq x,y$ there belongs a unique $s'$ and ${\bf w}$ which characterize it's position $s'$ on the field line ${\bf w}$. One can write

\begin{equation} gF^{\mu \nu} (z(s',{\bf w})) =  P \left\{gF^{\mu \nu} (z(s',{\bf w})) \exp[ig \iw Adz ]\right\} + O(g^2). \end{equation}
We have chosen the symbol $s'$ in $z(s',{\bf w})$ in order not to mix it up with the curve parameter $s$ in the path ordered exponential. The color matrix $F^{\mu \nu} (z(s',{\bf w}))$ has to be included in the path ordering prescription with respect to $s$ of the field line characterized by ${\bf w}$. In this way (3.7) behaves correctly under gauge transformations. 

Finally one has to multiply by $u(z;x,y)$ and perform the integration over $d^4z = \rho({\bf w}) ds'd^3w$, resulting in

\begin{eqnarray}\lefteqn {\int u(z;x,y) gF^{\mu \nu} (z) d^4z = } \nonumber\\
& & \int \rho ({\bf w}) u(z(s',{\bf w})) P \left\{gF^{\mu \nu} (z(s',{\bf w})) \exp[ig \iw Adz ]\right\} ds'd^3w + O(g^2). \end{eqnarray}  
This type of integral is a generalization
of the operator insertions into a Wilson loop introduced by Eichten
and Feinberg \cite{EF} and later on widely used in the literature. It is, however, more general, because the insertions are not equally distributed along the path, but weighted by the $s'$-dependence of $u(z(s',{\bf w}))$.

We summarize our representation for the quark propagator: It has the
decomposition (2.3) with

\newpage

\begin{eqnarray} s & = & m\Delta (x-y) \int \rho ({\bf w})P \exp[ig\iw
Adz ]d^3 w,\\
p & = & 0,  \\
v^\mu & = & \frac{i}{2} D^\mu (x) \left\{ \Delta
(x-y) \int \rho ({\bf w})P \exp[ig\iw Adz ]d^3
w\right\} \nonumber\\
& & - \frac{i}{2} \left\{ \Delta (x-y) \int \rho ({\bf w})P \exp[ig\iw
Adz ]d^3 w\right\}\stackrel{\leftarrow}{D}^{*\mu
}(y) \\
& &  - \frac{1}{2} D_\nu(x) \int \rho ({\bf w}) u(z(s',{\bf w})) P \left\{gF^{\mu \nu}(z(s',{\bf w}))  \exp[ig \iw Adz ]\right\}ds'd^3w \nonumber\\
& & - \frac{1}{2}  \int \rho ({\bf w})u(z(s',{\bf w})) P \left\{gF^{\mu \nu}(z(s',{\bf w}))  \exp[ig \iw Adz ]\right\}ds'd^3w \stackrel{\leftarrow}{D}^*_\nu(y), \nonumber\\
a^\mu & = & \frac{i}{4} \epsilon ^{\mu \nu \lambda
\kappa } \left\{D_\lambda (x) \int \rho ({\bf w})u(z(s',{\bf w})) P \left\{gF_{\nu \kappa }(z(s',{\bf w})) \exp[ig \iw Adz ]\right\}ds'd^3w \right.\nonumber\\
& & \left.- \int \rho ({\bf w})u(z(s',{\bf w})) P \left\{ gF_{\nu \kappa }(z(s',{\bf w})) \exp[ig \iw Adz ]\right\} ds'd^3w \stackrel{\leftarrow}{D}_\lambda ^*(y)\right\}, \\
t^{\mu \nu } & = & -\frac{m}{2}  \int \rho ({\bf w})u(z(s',{\bf w})) P \left\{
gF^{\mu \nu }(z(s',{\bf w}))P \exp[ig \iw Adz ]\right\} ds'd^3w.  \end{eqnarray}
We could have simplified the curly brackets $\{ \cdots \}$ in $v^\mu $ by $s/m$, but we left it in the present form in order to keep it applicable also in the case of small or vanishing mass. Obviously the last two lines in $v^\mu $ and the terms $a^\mu $ and $t^{\mu \nu }$ contain the same types of insertions into path ordered integrals.

Clearly the representation (2.3), (3.9)-(3.13) fulfills all properties
mentioned at the end of sect. 1.

\setcounter{equation}{0}\addtocounter{saveeqn}{1}%

\section{Weight function, stream lines, and a simple application}

It is now appropriate to switch to Euclidean space, i.e. put $x^0 = -ix_4^E,x^n = x_n^E, \gamma ^0 = \gamma _4^E, \gamma ^n = i\gamma _n^E$. In the following the index $E$ will be written explicitly only where it appears appropriate.
For the explicit calculations we make use of the rotation symmetry around the vector $(x-y)_\mu $. Choose, just for intermediate simplification of
notation, a system where $(x-y)_\mu $ has a four-component only and where ${\bf x} = {\bf y}= 0$. It is then
convenient to describe the vector $z_\mu $ by it's four-component $z_4$,
and ordinary three-dimensional polar coordinates $r,\Theta ,
\varphi $, i. e. 

 \begin{equation} z_\mu =(r \sin\Theta \cos \varphi ,r \sin \Theta
\sin \varphi ,r \cos \Theta , z_4).  \end{equation}
Let us now choose an appropriate parametrization of the various stream
lines. We classify them by the orthogonal distance $w$ of the line
from the midpoint $(x+y)/2$ between the sources, and by the angles $\Theta$ and
$\varphi $, thus $d^3w=dw d\Theta d\varphi$. Fig. 1 shows a stream line together with the parameters
introduced above. \\[1 ex]

\setlength{\unitlength}{2mm}
\begin{picture}(120,30)(0,0)
\put(11,0){\epsfig{file=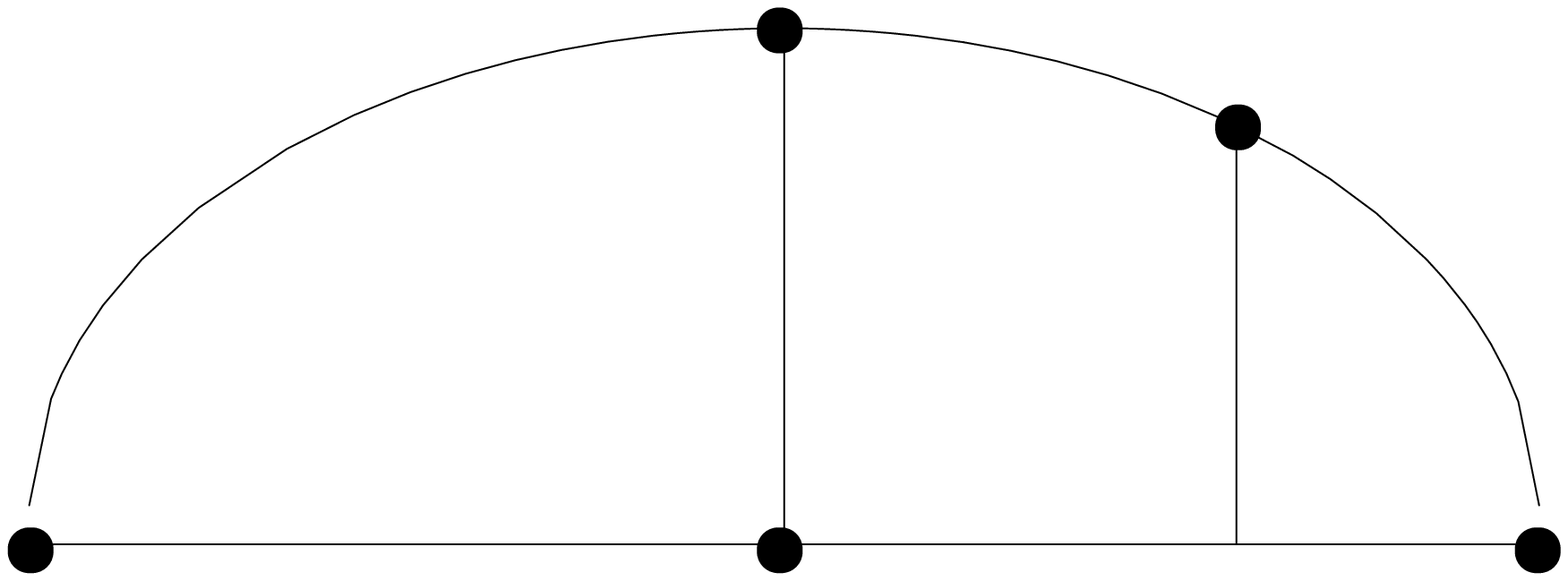,height=30\unitlength}}
\put(11.1,9.3){$y_\mu $}
\put(32,9.3){$(x+y)_\mu /2$}
\put(42.6,14){$ z_4 \longrightarrow$}
\put(58.9,9.3){$x_\mu $}
\put(49,27){$z_\mu $}
\put(36.9,19){$w$}
\put(51,19){$r$}
\put(36.5,25.5){$\zeta _\mu $}
\put(32.5,29){$\leftarrow s \rightarrow$}
\end{picture}

{\bf Fig. 1: } {\it The plane spanned by the stream line, and the coordinates introduced in the text (in the system where $x-y$ has a four-component only).}\\[1 ex]

The weight function $\rho ({\bf w})$ defined in (3.3) becomes

\begin{equation} \rho (w,\Theta ) = \frac{\partial
(z_1,z_2,z_3,z_4)}{\partial (r,\Theta,\varphi ,z_4)} \;
\frac{\partial (r,\Theta ,\varphi ,z_4)}{\partial (s,w,\Theta,\varphi
)} = r^2 \sin \Theta \; \frac{\partial (r,z_4)}{\partial (s,w)}.
\end{equation}
It is now very convenient that $\rho $ depends on ${\bf w}$ only, but
not on the curve parameter $s$. Therefore we can evaluate it at a suitable point. We choose the
symmetry point $\zeta $ in the middle of the stream line with the coordinates
$r=w,z_4 = (x+y)_4/2$. At this point we obviously have $\partial r/\partial s
=0$ and $\partial r/\partial w =1$, and thus $\partial
(r,z_4)/\partial (s,w) = -\partial z_4/\partial s = -u_4$, where we used the definition of the stream lines in (3.2). This allows to write down the weight function in closed form. 

\begin{equation} \rho (w,\Theta ) = -w^2 \sin \Theta \; u_4(\zeta ;x,y), \end{equation}
with  

\begin{equation} \zeta _\mu =(w\sin \Theta \cos \varphi ,w\sin \Theta \sin \varphi, w \cos \Theta,(x+y)_4/2).  \end{equation}
Obviously $u_4(\zeta ;x,y)$ is independent of the angles $\Theta ,\varphi $. 

It is convenient to put

\begin{equation} \rho (w,\theta ) = \frac{\hat{\rho }(w)}{4\pi }\sin \Theta, \end{equation}
such that the weight function $\hat{\rho }(w)$ is normalized to 

\begin{equation} \int _0^\infty \hat{\rho }(w)dw = 1. \end{equation}
We first give the resulting formulae for the special case of vanishing quark
mass which may be of some general interest:

\begin{equation} \Delta ^{(0)}(x) = -\frac{1}{4\pi ^2 x^2},  \end{equation}

\begin{equation} u^{(0)}(z;x,y) = \frac{1}{16 \pi ^4 (x-z)^2 (z-y)^2},  \end{equation}

\begin{equation} u_\mu ^{(0)}(z;x,y) = -\frac{(x-y)^2}{2\pi ^2}\left\{\frac{(x-z)_\mu }{(x-z)^4 (z-y)^2} +
\frac{(z-y)_\mu }{(x-z)^2 (z-y)^4} \right\} , \end{equation}

\begin{equation} \hat{\rho ^{(0)}}(w) = \frac{2w^2  |x-y|^3}{\pi
 [(x-y)^2/4 + w^2]^3}.  \end{equation}
The maximum of $\hat{\rho }^{(0)}$ is at $w ^{(0)}_{max}=|x-y|/(2\sqrt{2})$.

We now come to the general massive case. The free scalar
propagator $\Delta $ then is

\begin{equation} \Delta (x) = -\frac{m}{4\pi ^2x}K_1(mx),
\end{equation}
which gives

\begin{equation} u(z;x,y) = \frac{m^2}{16 \pi ^4} \frac{K_1(m|x-z|)K_1(m|z-y|)}{|x-z||z-y|}.  \end{equation}
With the relation $(K_1(z)/z)' = -K_2(z)/z$ for
the Kelvin function, one further obtains  

\begin{eqnarray} u_\mu (z;,x,y) = -\frac{m^2|x-y|}{4\pi ^2K_1(m|x-y|)}
\left\{ \frac{ K_2(m|x-z|)K_1(m|z-y|)}{(x-z)^2|z-y|}  (x-z)_\mu 
\right. \nonumber\\
\left. + \frac{ K_1(m|x-z|)K_2(m|z-y|)}{|x-z|(z-y)^2} (z-y)_\mu 
\right\}. \end{eqnarray}

\begin{equation} \hat{\rho }(w) = \frac{m^2w^2 (x-y)^2}{\pi }
\frac{K_1(m\sqrt{(x-y)^2/4+w^2})K_2(m\sqrt{(x-y)^2/4+w^2})}{K_1(m|x-y|)[(x-y)^2/4+w^2]^{3/2}}. \end{equation}
The weight function $\hat{\rho }(w)$ is trivially suppressed for small $w$ by the volume element in polar coordinates, it rises to a  maximum at some $w_{max}$, and decreases exponentially for large $w$.

In Fig. 2 we plot the function $\hat{\rho }(w)$ for fixed distance $|x-y|$ for various values of the mass. It is clearly seen how the maximum moves to the left if the quark mass increases.

\newpage

\setlength{\unitlength}{2mm}
\begin{picture}(120,30)(0,0)
\put(12,0){\epsfig{file=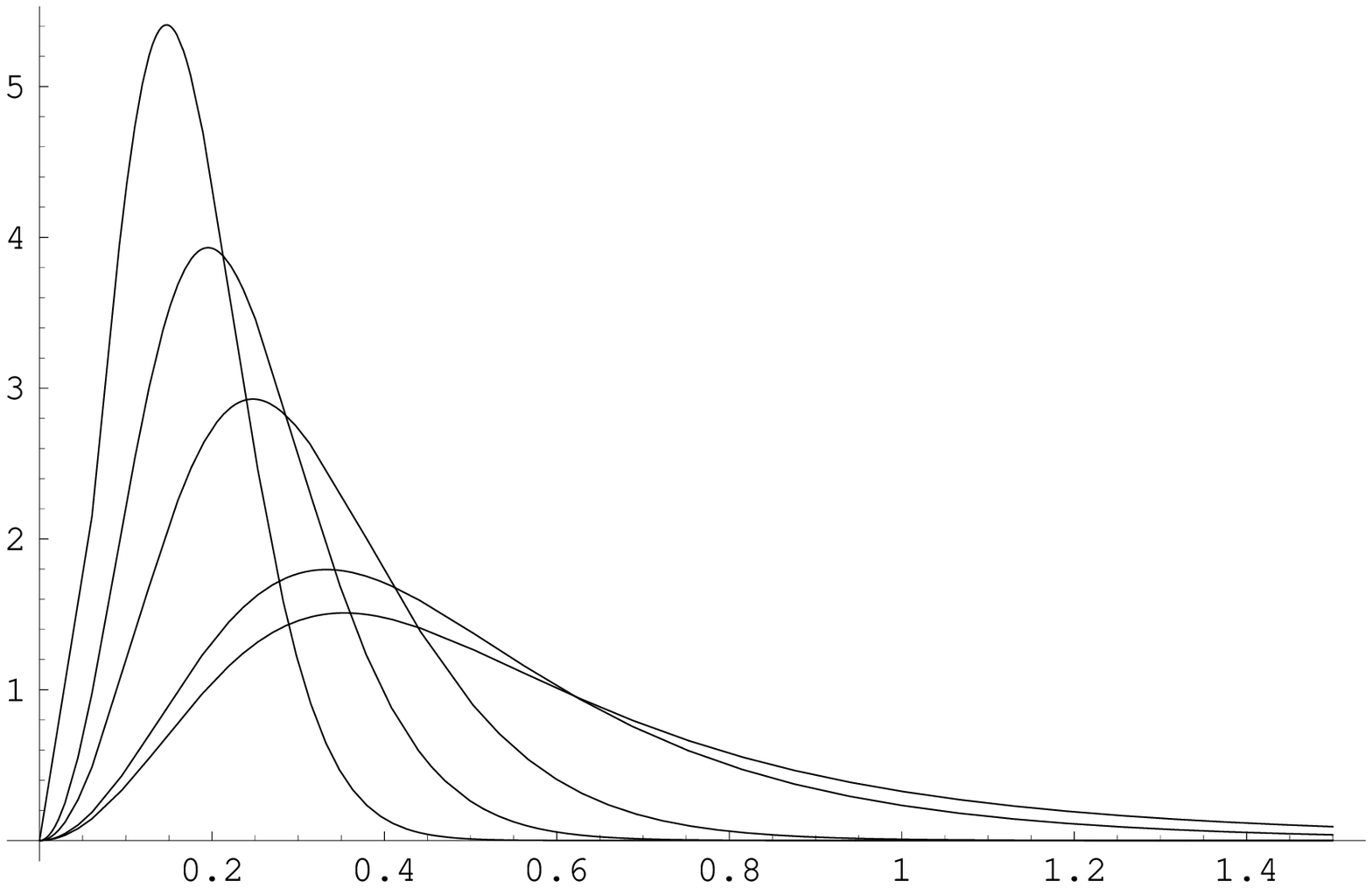,height=30\unitlength}}
\put(8,28){$\hat{\rho}(w)$}
\put(53,5){$w/|x-y|$}
\end{picture}

{\bf Fig. 2: } {\it The weight function $\hat{\rho }(w)$ defined in (4.6). The distance $|x-y|$ is fixed, we show the curves for the mass values $m|x-y| = 0,1,5,10,20$. The maximum moves from right to left and increases with increasing mass.} \\[1 ex]

In Fig. 3 we show the stream lines of the vector field $u_\mu $ for four values of $m|x-y|$.

\setlength{\unitlength}{1.3mm}
\begin{picture}(120,30)(0,0)
\put(0,0){\epsfig{file=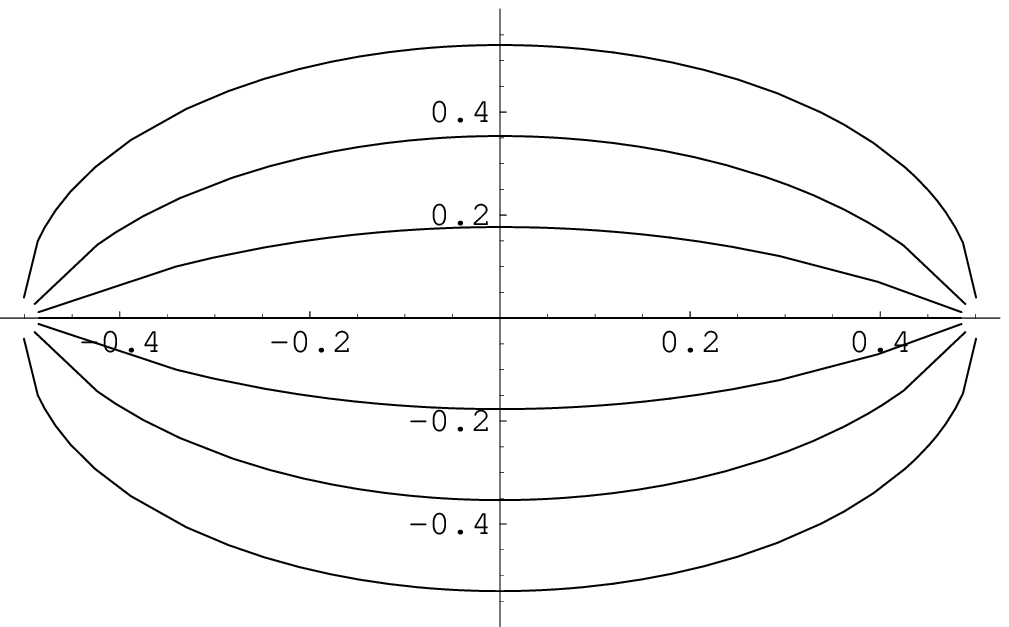,height=30\unitlength}}
\put(60,0){\epsfig{file=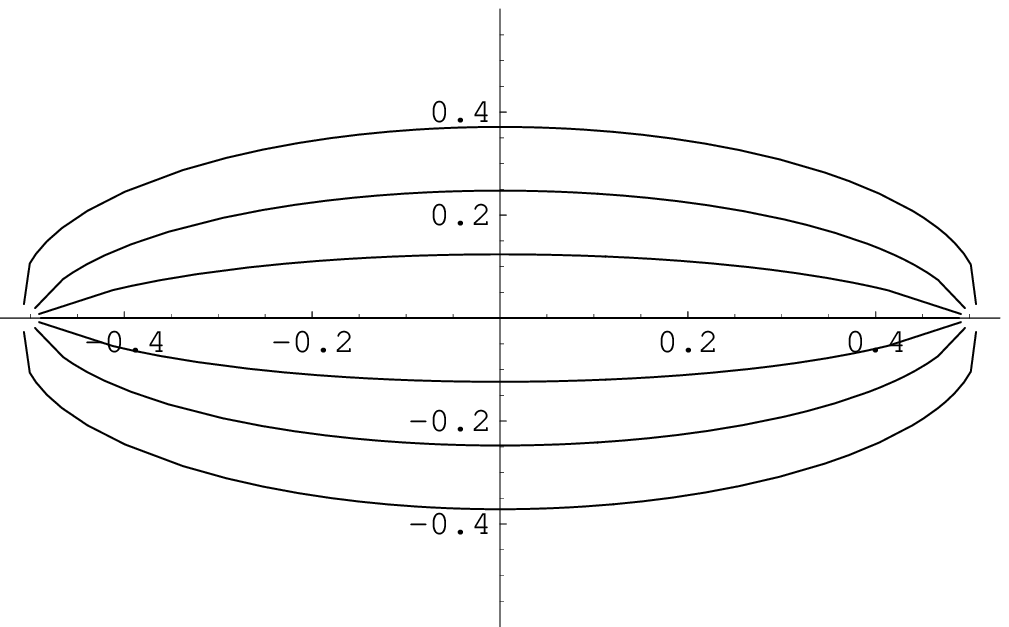,height=30\unitlength}}
\end{picture}

\setlength{\unitlength}{1.3mm}
\begin{picture}(120,30)(0,0)
\put(0,0){\epsfig{file=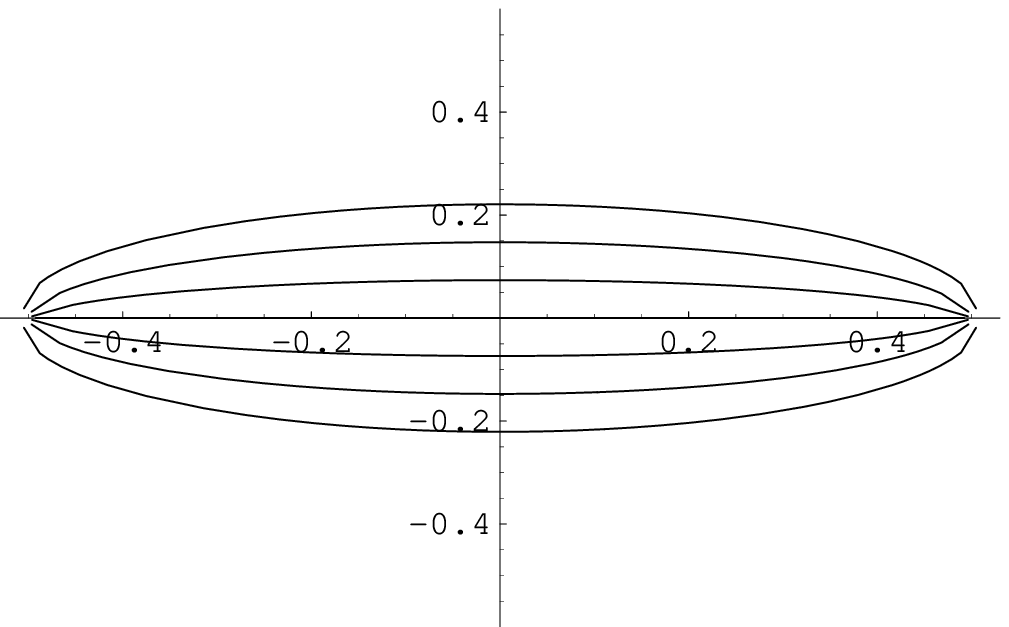,height=30\unitlength}}
\put(60,0){\epsfig{file=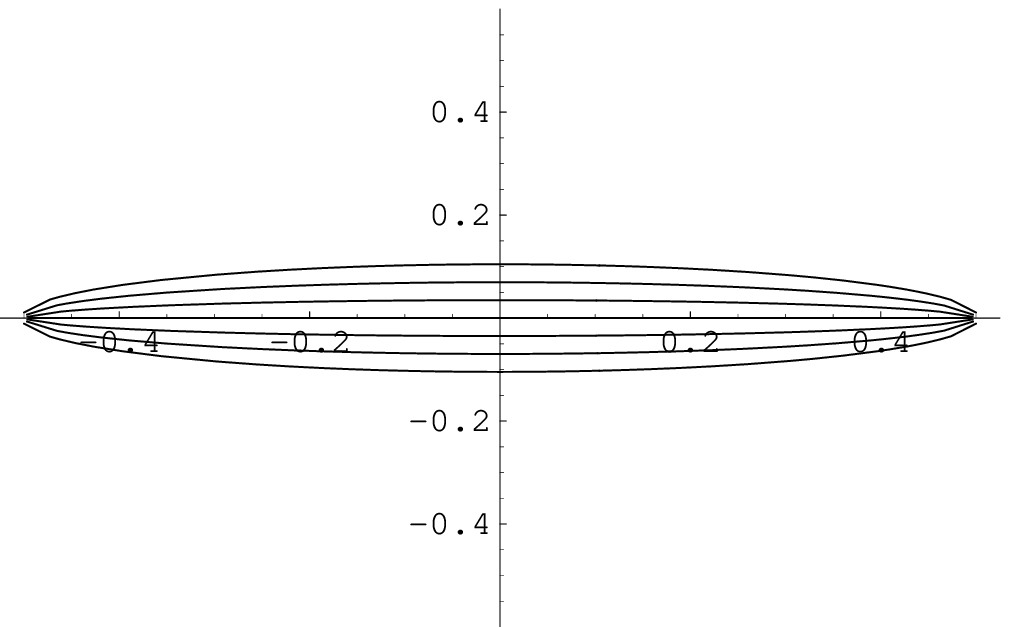,height=30\unitlength}}
\end{picture}

{\bf Fig. 3: } {\it The stream lines (3.2) for the cases (from left to right, top to bottom) a) $m|x-y| = 0$, b) $m|x-y| = 5$, c) $m|x-y| = 20$, d) $m|x-y| = 100$. The sources at $x$ and $y$ are located at $\pm 0.5 $. We show the stream lines for the values of $w/w_{max} = 0,\pm 0.5, \pm 1, \pm 1.5$.}\\[1 ex]

For vanishing mass the lines spread out widely in space, up to $w$ of the order of $|x-y|$. For increasing mass they concentrate more and more to the straight line connection. Apparently the product $m|x-y|$ has to become quite large, however, in order to get a sizeable concentration. 

The large mass limit will now be investigated analytically. We use the asymptotic behavior $K_n(z) \sim \sqrt{\pi /(2z)}e^{-z}$ for $z \rightarrow 
\infty $ and  get

\begin{equation} \Delta (x) \sim -\frac{1}{2} \frac{\sqrt{m}}{(2 \pi  x)^{3/2}}e^{-mx} \mbox{\quad for } m \rightarrow \infty,  \end{equation}

\begin{equation} u(z;x,y) \sim \frac{m}{32\pi ^3}\frac{e^{-m[|x-z| + |z-y|]}}{[|x-z| |z-y|]^{3/2}} \mbox{\quad for } m \rightarrow \infty,  \end{equation}

\begin{eqnarray} u_\mu (z;x,y) & \sim  & -\frac{1}{2} \left( \frac{m |x-y|}{2 \pi |x-z| |z-y|} \right) ^{3/2} \left\{ \frac{(x-z)_\mu }{|x-z|} + \frac{(z-y)_\mu}{|z-y|}\right\} e^{m[|x-y| - |x-z| - |z-y|]}\nonumber\\
& & \mbox{\quad for } m \rightarrow \infty,  \end{eqnarray}

\begin{equation} \hat{\rho }(w) \sim \frac{m^{3/2} w^2|x-y|^{5/2}}{\sqrt{2 \pi}[(x-y)^2/4 + w^2]^2}  \exp [m(|x-y| - 2\sqrt{(x-y)^2/4 + w^2})] \mbox{\quad for } m \rightarrow \infty.  \end{equation}
Obviously only $w$ of order $\sqrt{|x-y|/m}$ are now of relevance in the weight function. Therefore one may expand the square roots and gets the simple result

\begin{equation} \hat{\rho }(w) \sim {\frac{16}{\sqrt{2\pi }}}(\frac{m}{|x-y|})^{3/2} w^2 \exp [-\frac{2mw^2}{|x-y|}] \mbox{\quad for } m \rightarrow \infty.  \end{equation}
This is just the picture which one expects for large mass. Only stream lines near the straight line connection essentially contribute, the maximum of $\hat{\rho }(w)$ is at $w_{max} = \sqrt{|x-y|/(2m)}$. 

If the mass is large enough, such that the variation of the
gluon field in transversal direction becomes negligible, all line integrals
give the same contribution. The weighted superposition over the paths
can then simply be replaced by the path along the straight line connection. This means that one has effectively a three-dimensional $\delta $-function in
transversal direction. The situation looks now similar to the case of the 
static propagator  but with an important difference. While the
static propagator (1.8) singles out a special reference frame, our propagator is manifestly Lorentz covariant. It is the vector $(x-y)_\mu $
which specifies the direction of propagation.

This has a simple consequence. In the limit of large mass, the scalar function $s$ in (3.9) becomes

\begin{equation} s \sim -\frac{1}{2}(\frac{m}{2 \pi |x-y|})^{3/2} e^{-m|x-y|} P \exp [-ig \int _y^xAdz], \end{equation}
with the Euclidean path along the straight line connecting $x$ and $y$.

Consider now the term $s$ plus the first two lines of $v^\mu \gamma _\mu $ in (3.11). This sum can be written as

\begin{eqnarray} s & - & \frac{1}{2m}(D^E_\mu (x) s - s \stackrel{\leftarrow }{D}^{*E}_\mu (y)) \gamma ^E_\mu \nonumber\\
 \sim &  - & (\frac{m}{2 \pi |x-y|})^{3/2}\frac{1+ \gamma _\mu ^E(x-y)_\mu / |x-y|}{2} e^{-m|x-y|} P \exp [-ig \int _y^xA dz]. \end{eqnarray}
In contrast to the $e^{-m|x_4-y_4|}$ of the static approximation which falls off with the euclidean time difference, our $e^{-m |x-y|}$ falls off with the euclidean distance. This means that the Hamiltonian is correctly given by the full relativistic energy. Furthermore we also get the correct projection operator for the $\gamma $-matrices. 

We finally discuss the terms $a^\mu,t^{\mu \nu }$ in (3.12), (3.13) which can be treated rather simply. We specialize to the case ${\bf x} = {\bf y}$ and put $x_4 - y_4 = T>0$. In the large mass limit we can replace $\rho ({\bf w})$ by $\delta^{(3)}({\bf w})$. The $s'$-integrations can be written as follows:

\begin{equation} u(z(s',0))ds' = u(t) \frac{ds'}{dt} dt = \frac{u(t)}{u_4(t)} dt = - \frac{ e^{-mT}}{8\sqrt{2} \pi^{3/2}T^{3/2}\sqrt{m}} dt.   \end{equation}
In the second step we used the 4-component of (3.2), in the third step we introduced the asymptotic formulae (4.16), (4.17) for the special case where $z$ lies on the line connecting $x$ and $y$.  In  $a^\mu $ we only need to consider the index $\lambda = 4$ which gives a leading factor $(-m)$ from the differentiation of $e^{-mT}$. All other contributions are suppressed by higher powers of $1/m$. In $t^{\mu \nu }\sigma _{\mu \nu }$ only spatial indices $\mu = m,\nu = n$ survive if we concentrate on the diagonal part of the Dirac matrices. The transition to $2\times 2$-matrices than gives

\begin{eqnarray} 
\mbox{in the axial vector (3.12):\quad} a^\mu\gamma^5 \gamma _\mu & \Rightarrow &  \epsilon_{mn4k} F_{nk}\gamma ^5 \gamma _m  \Rightarrow \epsilon_{mnk}F_{nk}\sigma _m \Rightarrow  2 B_m\sigma _m,\nonumber\\
\mbox{in the tensor (3.13):\quad}   t^{\mu \nu} \sigma _{\mu \nu} & \Rightarrow & F_{mn} \sigma _{mn}  \Rightarrow  \epsilon _{mnk}F_{mn} \sigma _k \Rightarrow 2 B_k\sigma_k.
\end{eqnarray}
Both contributions give identical magnetic field insertions which add up. 

The resulting expressions have to be combined with the leading term (4.21) where we can drop the projector $(1+\gamma _4^E)/2$. This gives a spin dependent expression of the form

\begin{equation} -(\frac{m}{2\pi T})^{3/2}e^{-mT} P\exp[-ig\int _y^xAdz](1+\frac{g}{m}\int{\bf B}(t) {\bf s}dt). \end{equation}
If we take the product of the quark- and the antiquark propagator which arises in the four-point Green function, focus on the product term of the two magnetic field insertions, and extract the Hamiltonion in the usual way \cite{EF},\cite{rev}, we immediately obtain the spin spin and the tensor terms with the correct representations for the corresponding potentials $V_4$ and $V_3$. In the static approach these terms only arise as higher order corrections. 

Spin orbit terms are momentum dependent and therefore involve moving quarks. These are, of course, contained in our formalism, but the derivation is slightly more complicated. Spin independent corrections are obtained even harder. Therefore we will not discuss those in this first application, but be content with the simple and correct derivation of spin spin and tensor terms given above.

\setcounter{equation}{0}\addtocounter{saveeqn}{1}%

\section{Conclusions and outlook}

Let us first compare our representation
(2.3), (3.9)-(3.13) with the static approximation (1.8). The static
approximation is, drastically speaking, completely wrong everywhere.
It is completely wrong for ${\bf x} \neq {\bf y}$ where it vanishes, but it is also completely wrong for ${\bf x} = {\bf y}$ because
it has a $\delta $-function there which is not present in the exact
propagator. These features survive if one treats the neglected spatial
part $i\gamma ^mD_m$ in the field equation as perturbation. In any
finite order of this perturbation the approximated propagator
vanishes for ${\bf x} \neq {\bf y}$, while  higher order derivatives of $\delta ^{(3)} ({\bf x} - {\bf y})$ appear. The miracle that one can nevertheless obtain useful results from this propagator is due to the fact that the perturbation series turns out to become an expansion with respect to $1/m$. As long as $<p^2/m^2>$ is small, the results are reliable. 

The propagator proposed in the present work is manifestly covariant and appears to have a reasonable structure everywhere.
This advantage is payed by a more complicated form,
which, however, looks very natural physically. Not only one path
ordered integral, but a whole set of them contribute. We saw how the
paths near the straight line connection dominate for large mass. We believe that it will also be possible to get useful
information for finite mass. For an investigation of the quark-antiquark interaction one should start, as usual, with the gauge invariant $q\bar{q}$ four-point Green function and integrate out the fermion fields. Instead of the familiar Wegner-Wilson loop one will now obtain a superposition of loops where the straight paths in time direction are replaced by the stream lines making up our propagator. Quite a lot of knowledge how to treat such loops has been accumulated by various authors which can be used for this investigation.

A comparison with the Feynman-Schwinger representation (see e.g. \cite{schwi}) is also instructive. This representation of the propagator is formally exact and has essentially the form of a quantum mechanical Green function. It can therefore be written as a path integral. In the literature \cite{apath} one also finds approximate path integral representations for the propagator and the quark-antiquark kernel valid up to order $1/m^2$. In both cases one has, as usual in this formalism, a sum over {\em all paths}, which is, conceptionally as well as technically, a rather delicate concept. Even for rather simple situations the path integral cannot be evaluated. In our case, on the other hand, we have only line integrals along {\em a well defined set of field lines}. This is a much simpler situation. Our representation stands just between approximations which involve a single path only and those requiring a sum over all paths.

Besides the application of the present propagator for large as well as for finite mass, there is another topic which  should be worked out. This is the systematic improvement of our propagator. We don't have a simple differential equation for it, as it is available in the case of the static propagator. Therefore one probably has to improve higher orders of perturbation theory directly and transform them into gauge covariant expressions in an analogous way as done here for the first order.
We emphasize, however, that such an improvement does not appear necessary for many purposes. The present form gives already the correct relativistic
energy of a free particle together with the correct spin projectors. Furthermore we have seen that it gives the correct spin-spin and tensor forces for heavy quark-antiquark systems. To get these ''relativistic corrections''  from the static propagator one has to make quite complicated manipulations. We expect that all other relativistic corrections can also be obtained with some more effort.

Therefore there are good reasons to believe that the suggested expression for the propagator will turn out quite useful already in it's present form.\\ [1 ex]

{\bf Acknowledgement:} I thank H. G. Dosch for reading the manuscript.

\end{document}